\documentclass[twocolumn,epjc3]{svjour3mod}

\usepackage[colorlinks,citecolor=blue,urlcolor=blue,linkcolor=blue]{hyperref}
\usepackage{flushend}
\usepackage{graphicx,colordvi}
\usepackage{amsmath,amssymb,slashed,cite}
\usepackage{booktabs,tabulary}
\usepackage{dsfont}
\usepackage{color}
\usepackage{cite}
\usepackage{siunitx}

\usepackage{ulem}

\allowdisplaybreaks[1]

\newcommand{\MeV}{\,\text{MeV}}
\newcommand{\GeV}{\,\text{GeV}}

\newcommand{\beq}{\begin{equation}}
\newcommand{\eeq}{\end{equation}}
\newcommand{\diff}{\text{d}}

\newcommand{\Mw}{M_\omega}
\newcommand{\Mphi}{M_\phi}
\newcommand{\barMw}{\bar M_\omega}
\newcommand{\barMphi}{\bar M_\phi}

\newcommand{\gwg}{g_{\omega\gamma}}
\newcommand{\Gw}{\Gamma_\omega}
\newcommand{\Gphi}{\Gamma_\phi}
\newcommand{\barGw}{\bar\Gamma_\omega}
\newcommand{\barGphi}{\bar\Gamma_\phi}

\newcommand{\eps}{\epsilon}

\newcommand{\siv}{s_\text{iv}}

\newcommand{\bsp}{\begin{sloppypar}}
\newcommand{\esp}{\end{sloppypar}}

\begin{document}

\title{Hadronic vacuum polarization and vector-meson\\[2mm] resonance parameters from $\boldsymbol{e^+e^-\to\pi^0\gamma}$}
\author{Bai-Long Hoid\thanksref{addr2}
\and
Martin Hoferichter\thanksref{addr1}
        \and
        Bastian Kubis\thanksref{addr2} 
}
\institute{Helmholtz-Institut f\"ur Strahlen- und Kernphysik (Theorie) and
Bethe Center for Theoretical Physics, Universit\"at Bonn, 53115 Bonn, Germany\label{addr2}
\and 
Albert Einstein Center for Fundamental Physics, Institute for Theoretical Physics, University of Bern, Sidlerstrasse 5, 3012 Bern, Switzerland\label{addr1}
}

\date{}

\maketitle

\begin{abstract}
\bsp
We study the reaction $e^+e^-\to\pi^0\gamma$ based on a dispersive representation of the underlying $\pi^0\to\gamma\gamma^*$ transition form factor. As a first application, we evaluate the contribution of the $\pi^0\gamma$ channel to the hadronic-vacuum-polarization correction to the anomalous magnetic moment of the muon. We find $a_\mu^{\pi^0\gamma}\big|_{\leq 1.35\GeV}=43.8(6)\times 10^{-11}$, in line with evaluations from the direct integration of the data. Second, our fit determines the resonance parameters of $\omega$ and $\phi$. We observe good agreement with the $e^+e^-\to3\pi$ channel, explaining a previous tension in the $\omega$ mass between $\pi^0\gamma$ and $3\pi$ by an unphysical phase in the fit function. Combining both channels we find 
$\barMw=782.736(24)\MeV$ and $\barMphi=1019.457(20)\MeV$ for the masses including vacuum-polarization corrections. The $\phi$ mass agrees perfectly with the PDG average, which is dominated by  determinations from the $\bar K K$ channel, demonstrating consistency with $3\pi$ and $\pi^0\gamma$. For the $\omega$ mass, our result is consistent but more precise, exacerbating tensions with the $\omega$ mass extracted via isospin-breaking effects from the $2\pi$ channel.   
\esp
\end{abstract}

%-----------------------------------------------------------------------------
\section{Introduction}
\label{sec:introduction}
%-----------------------------------------------------------------------------

\bsp
The vector mesons $\omega$ and $\phi$ are narrow states compared to other hadronic resonances in the low-energy QCD spectrum. In the case of the $\omega$, this is because two-body decays are either forbidden by $G$ parity ($2\pi$) or require electromagnetic interactions ($\pi^0\gamma$, $\eta\gamma$), so that the dominant decay proceeds into $3\pi$. In contrast, for the $\phi$ a $G$-parity conserving two-body decay into $\bar K K$ is possible, but suppressed by  
very small phase space, while the decay into $3\pi$ is small due to the Okubo--Zweig--Iizuka rule~\cite{Okubo:1963fa,Zweig:1964jf,Iizuka:1966fk}.
Accordingly, the most precise information on the mass of the $\phi$ comes from $e^+e^-\to \bar K K$~\cite{Achasov:2000am,Akhmetshin:2003zn,Lees:2013gzt,Lees:2014xsh,Kozyrev:2017agm}, which indeed dominates the PDG average~\cite{PDG:2020}. For the determination of the $\omega$ mass, the reaction $e^+e^-\to 3\pi$ is the primary source of information~\cite{Akhmetshin:2003zn,Achasov:2003ir}, but here the three-particle nature of the decay complicates a reliable extraction of the resonance parameters. In particular, there is a significant tension with the mass determination from $e^+e^-\to\pi^0\gamma$~\cite{Akhmetshin:2004gw}, which together with $\bar p p\to\omega\pi^0\pi^0$~\cite{Amsler:1993pr} leads to a scale factor $S=1.9$ in the PDG average. In this work, we consider the reaction $e^+e^-\to\pi^0\gamma$ using a dispersive representation of the $\pi^0\to\gamma\gamma^*$ transition form factor (TFF), which together with our previous work on the $3\pi$ channel~\cite{Hoferichter:2019gzf} allows us to present a combined determination of the $\omega$ and $\phi$ resonance parameters within the same framework consistent with the constraints from analyticity, unitarity, and crossing symmetry as well as low-energy theorems.  

These constraints, as incorporated in the dispersive representation of the TFF~\cite{Hoferichter:2014vra,Hoferichter:2018dmo,Hoferichter:2018kwz}, are not only valuable for a reliable extraction of resonance parameters, but also define a global fit function for the cross section that allows one to check the consistency of the data sets with these general principles. Applications to the $e^+e^-\to2\pi$~\cite{Colangelo:2018mtw,Ananthanarayan:2018nyx,Davier:2019can,Colangelo:2020lcg} and $e^+e^-\to3\pi$~\cite{Hoferichter:2019gzf}  channels have provided such analyses for the two dominant channels in the hadronic-vacuum-polarization (HVP) contribution to the anomalous magnetic moment of the muon $a_\mu$. Here, we will study the $e^+e^-\to\pi^0\gamma$ channel in the same spirit. Since the total contribution is about an order of magnitude smaller than the one of the $3\pi$ channel, very large relative changes would be required to notably influence the Standard Model prediction $a_\mu^\text{SM}=116\,591\,810(43)\times 10^{-11}$~\cite{Aoyama:2020ynm,Aoyama:2012wk,Aoyama:2019ryr,Czarnecki:2002nt,Gnendiger:2013pva,Davier:2017zfy,Keshavarzi:2018mgv,Colangelo:2018mtw,Hoferichter:2019gzf,Davier:2019can,Keshavarzi:2019abf,Kurz:2014wya,Melnikov:2003xd,Masjuan:2017tvw,Hoferichter:2018dmo,Hoferichter:2018kwz,Gerardin:2019vio,Colangelo:2017qdm,Colangelo:2017fiz,Bijnens:2019ghy,Colangelo:2019lpu,Colangelo:2019uex,Blum:2019ugy,Colangelo:2014qya} and thus the tension with the BNL measurement $a_\mu^\text{exp}=116\,592\,089(63)\times 10^{-11}$~\cite{Bennett:2006fi}. 
However, in view of recent results from lattice QCD~\cite{Borsanyi:2020mff} that suggest large modifications of the hadronic cross section at low energies~\cite{Crivellin:2020zul,Keshavarzi:2020bfy}, any further corroboration of the phenomenological HVP evaluation, especially for the channels relevant below $1\GeV$ such as $\pi^0\gamma$, is certainly worthwhile---in anticipation of improved measurements at Fermilab~\cite{Grange:2015fou} and J-PARC~\cite{Abe:2019thb}.                 
\esp 

The paper is organized as follows: in Sect.~\ref{sec:TFF} we review the dispersive formalism for the pion TFF and the $e^+e^-\to\pi^0\gamma$ cross section, which is then applied in Sect.~\ref{sec:fits} to fit the available data sets. In Sect.~\ref{sec:amu} we discuss the consequences for the HVP contribution to $a_\mu$, in Sect.~\ref{sec:resonances} the combined analysis of the $\omega$ and $\phi$ resonance parameters from $e^+e^-\to 3\pi$ and $e^+e^-\to\pi^0\gamma$. We close with a summary in Sect.~\ref{sec:summary}.

%-----------------------------------------------------------------------------
\section{Time-like pion transition form factor and $\boldsymbol{e^+e^-\to\pi^0\gamma}$ cross section}
\label{sec:TFF}
%-----------------------------------------------------------------------------
\bsp
Based on the unitarity relation and its crucial building blocks, a once-subtracted dispersive representation for the time-like singly-virtual TFF $F_{\pi^0\gamma^*\gamma^*}(q^2,0)$ was constructed in~\cite{Hoferichter:2014vra},  
\begin{align}
\label{eq:pigamma-os_TFF_tl}
F_{\pi^0\gamma^*\gamma^*}&(q^2,0)=F_{\pi\gamma\gamma}
+\frac{1}{12\pi^2}\int_{4M_\pi^2}^\infty \diff s' \frac{q_\pi^3(s')(F_\pi^{V}(s'))^* }{s'^{3/2}} \notag\\
\times &\bigg\{f_1(s',q^2)-f_1(s',0) 
+\frac{q^2}{s'-q^2}f_1(s',0)\bigg\},
\end{align}
where $q_\pi(s)=\sqrt{s/4-M_\pi^2}$,  $F_\pi^{V}(s)$ is the pion vector form factor, and $f_1(s,q^2)$ is the partial-wave amplitude for $\gamma^*\to3\pi$~\cite{Niecknig:2012sj,Hoferichter:2012pm,Hoferichter:2014vra,Hoferichter:2017ftn}, as a generalization of previous studies of the $\omega/\phi\to\pi^0\gamma^*$ TFFs~\cite{Schneider:2012ez,Danilkin:2014cra}.  
In particular, $F_{\pi^0\gamma^*\gamma^*}(q^2,0)$ was studied in~\cite{Hoferichter:2014vra} as a first step towards the doubly-virtual
space-like TFF~\cite{Hoferichter:2018dmo,Hoferichter:2018kwz}, which determines the strength of the pion-pole contribution in a dispersive approach to hadronic light-by-light scattering~\cite{Hoferichter:2013ama,Colangelo:2014dfa,Colangelo:2014pva,Colangelo:2015ama}, to demonstrate the consistency between $3\pi$ and $\pi^0\gamma$ data. Similarly, the $\omega$ and $\phi$ TFFs become relevant for the description of the left-hand cuts in the two-pion contributions~\cite{GarciaMartin:2010cw,Hoferichter:2011wk,Moussallam:2013una,Danilkin:2018qfn,Hoferichter:2019nlq,Danilkin:2019opj}. 

$F_{\pi\gamma\gamma}$ denotes the normalization at $q^2=0$, as determined at leading order by the Wess--Zumino--Witten anomaly~\cite{Wess:1971yu,Witten:1983tw}  
\beq
\label{Fpigg}
F_{\pi\gamma\gamma}=\frac{1}{4\pi^2 F_\pi}=0.2745(3)\GeV^{-1}.
\eeq
This value, obtained from the pion decay constant $F_\pi=92.28(10)\MeV$~\cite{PDG:2020}, agrees with the recent PrimEx-II measurement of the neutral-pion life time~\cite{Larin:2020}, which implies $F_{\pi\gamma\gamma}=0.2754(21)\GeV^{-1}$. 
The relation between the $e^+e^- \to \pi^0 \gamma$ cross section and the pion TFF, calculated from the dispersion relation~\eqref{eq:pigamma-os_TFF_tl}, 
reads
\begin{align}
\label{eq:pigamma-cross-sec-pig}
\sigma^0_{e^+e^- \to \pi^0 \gamma}(q^2) =\frac{2\pi^2\alpha^3}{3} \frac{(q^2-M_{\pi^0}^2)^3}{q^6} \big|F_{\pi^0\gamma^*\gamma^*}(q^2,0)\big|^2,
\end{align}
where $\alpha=e^2/(4\pi)$ and we neglected the mass of the electron. Strictly speaking, the dispersion relation~\eqref{eq:pigamma-os_TFF_tl} applies to the pure QCD process without further radiative correction, so that~\eqref{eq:pigamma-cross-sec-pig} describes the bare cross section $\sigma^0_{e^+e^- \to \pi^0 \gamma}(q^2)$ excluding vacuum-polarization (VP) corrections. Accordingly, the mass parameters for $\omega$ and $\phi$ extracted from the fit do not include these VP corrections, in contrast to the PDG convention, see Sect.~\ref{sec:resonances}. We use the VP routine from~\cite{Keshavarzi:2018mgv} to remove VP from the experimental cross sections. 

The isoscalar contribution, corresponding to $f_1(s',q^2)-f_1(s',0) $ in the integrand of~\eqref{eq:pigamma-os_TFF_tl}, was calculated in~\cite{Hoferichter:2014vra} using the previously determined partial wave $f_1(s,q^2)$, where the normalization function $a(q^2)$ was fixed from a fit to $e^+e^-\to3\pi$ data;  the isovector part, the last term in~\eqref{eq:pigamma-os_TFF_tl},  was determined using a finite matching point of $\SI{1.2}{\GeV}$ and a normalization at $q^2=0$ fixed to the chiral anomaly $F_{3\pi}$ for the $\gamma\to3\pi$ amplitude~\cite{Adler:1971nq,Terentev:1971cso,Aviv:1971hq}. We will implement the same constraint here, i.e., including quark-mass corrections~\cite{Hoferichter:2012pm,Bijnens:1989ff}  
\beq
\label{F3pi}
a(0)=\frac{F_{3\pi}}{3}\times 1.066(10),\qquad F_{3\pi}=\frac{1}{4\pi^2F_\pi^3}.
\eeq 
We stress that in contrast to $F_{\pi\gamma\gamma}$, whose anomaly-constraint~\eqref{Fpigg} has been confirmed by PrimEx-II at the level of $0.8\%$, the chiral prediction for $F_{3\pi}$ has only been tested experimentally with $10\%$ precision, from Primakoff measurements~\cite{Antipov:1986tp}
and $\pi^- e^-\to \pi^- e^-\pi^0$~\cite{Giller:2005uy}. In the remainder of this paper, we assume that $F_{3\pi}$ follows the $F_{\pi\gamma\gamma}$ precedent, so that the remaining uncertainty in~\eqref{F3pi}, from the quark-mass renormalization, becomes subleading compared to other sources of systematic uncertainty in the dispersive representation of the TFF. In view of open questions regarding the role of subleading terms in the chiral expansion of the $\pi^0\to\gamma\gamma$ amplitude~\cite{Bijnens:1988kx,Goity:2002nn,Ananthanarayan:2002kj,Kampf:2009tk,Gerardin:2019vio}, a more stringent test of $F_{3\pi}$ would be highly desirable, which could be achieved with data on $\gamma\pi^-\to \pi^-\pi^0$ taken in the COMPASS Primakoff program~\cite{Seyfried}, using the dispersive framework proposed in~\cite{Hoferichter:2012pm,Hoferichter:2017ftn}. 

\begin{table*}[t]
	\centering
	\renewcommand{\arraystretch}{1.3}
	\small
	\begin{tabular}{l c c c}
		\toprule
		Experiment & Region of $\sqrt{s}$ [GeV] & \# data points & Normalization uncertainty\\
		\midrule
		SND 2000~\cite{Achasov:2000zd} & $[0.99,1.03]$ & $12$ & $3.3\%$ \\
		SND 2003~\cite{Achasov:2003ed} & $[0.60,0.97]$ & $30$ & all systematics\ \\
		SND 2016~\cite{Achasov:2016bfr} & $[0.63,1.35]$ & $60$ & all systematics\\
		SND 2018~\cite{Achasov:2018ujw} & $[1.08,1.35]$ & $5$  &  all systematics\\\midrule
		CMD-2 2005~\cite{Akhmetshin:2004gw} & $[0.60,1.31]$ & $46$ & $6.0\%$\\
		\bottomrule
		\renewcommand{\arraystretch}{1.0}
	\end{tabular}
	\caption{Summary of the $e^+e^-\to\pi^0\gamma$ data sets. For~\cite{Achasov:2018ujw} only data points for $\sqrt{s}<\SI{1.4}{\GeV}$ are included, as the cross section in the  region $(1.4\text{--}2.0)\,\SI{}{\GeV}$ was found to be consistent with zero. In the last column we indicate the size of the systematic errors that we interpret as a normalization-type uncertainty and therefore assume to be $100\%$ correlated.}
	\label{tab:pigamma-data_sets}
\end{table*}

As already remarked in~\cite{Hoferichter:2014vra}, the normalization function  $a(q^2)$ could also be determined by a fit to $e^+e^-\to \pi^0\gamma$ instead of the $3\pi$ channel. We follow this approach in the present work and consider an update of this once-subtracted analysis based on the improved parameterization for $a(q^2)$ developed in~\cite{Hoferichter:2018dmo,Hoferichter:2018kwz}, including a conformal polynomial to be able to describe the inelastic effects that were found to be relevant in $e^+e^-\to 3\pi$ above the $\phi$ resonance~\cite{Hoferichter:2019gzf}. For the details of the calculation of $f_1(s,q^2)$ we refer to~\cite{Hoferichter:2018dmo,Hoferichter:2018kwz,Hoferichter:2019gzf}, but reiterate the free parameters that enter the dispersive representation for the normalization function $a(q^2)$: apart from the $\omega$ and $\phi$ resonance parameters, these are their residues $c_\omega$ and $c_\phi$, as well as, potentially, further free parameters in the conformal polynomial. For the evaluation of the final dispersion relation~\eqref{eq:pigamma-os_TFF_tl}, we choose an integration cutoff $\siv$ above which an asymptotic behavior $\sim 1/s$ is assumed for both $F_\pi^{V}(s)$ and $f_1(s,q^2)$~\cite{Froissart:1961ux,Martin:1962rt, Farrar:1979aw, Duncan:1979hi,Efremov:1979qk}. The isovector part is updated as well in line with the isoscalar contribution.   

The systematic uncertainties of the dispersive representation are taken into account as follows: the pion vector form factor $F_\pi^{V}(s)$ is calculated with different variations of the Omn\`es function~\cite{Omnes:1958hv} using different phase shifts~\cite{GarciaMartin:2011cn,Caprini:2011ky} as in~\cite{Hoferichter:2018kwz}; in the meantime, the integration cutoffs $\Lambda_{3\pi}$ in the solution of the $\gamma^*\to3\pi$ Khuri--Treiman equations~\cite{Khuri:1960zz}   and $\sqrt{\siv}$ in the solution of the pion TFF~\eqref{eq:pigamma-os_TFF_tl} are varied in the range $(1.8\text{--}2.5)\,\SI{}{\GeV}$; lastly, the asymptotic behavior of the imaginary part of the conformal polynomial is varied as in~\cite{Hoferichter:2019gzf}.  The central values of the cross sections are obtained by the best fits to the data sets scanning over the variations of these quantities.  The systematic uncertainties are defined as the maximum deviations of all the variations from the central cross sections. 
\esp

%-----------------------------------------------------------------------------
\section{Fits to $\boldsymbol{e^+e^-\to\pi^0\gamma}$ data}
\label{sec:fits}
%-----------------------------------------------------------------------------
%--------------------------------------------------------------------------------------------------------
\subsection{Data sets and normalization uncertainties}
\label{sec:pigamma-data_sets}
%--------------------------------------------------------------------------------------------------------

In addition to the $e^+e^-\to\pi^0\gamma$ cross section measurements~\cite{Achasov:2000zd,Achasov:2003ed,Akhmetshin:2004gw} already included in~\cite{Hoferichter:2014vra}, there are two new data sets, the most accurate new data determined from the whole data sample of the SND experiment~\cite{Achasov:2016bfr} and another one that explored a new region between  $1.4$ and $\SI{2.0}{\GeV}$~\cite{Achasov:2018ujw}.  The full data sets that we consider in our analysis are listed in Table~\ref{tab:pigamma-data_sets}.  These measurements were performed  at the  VEPP-2M collider with the SND~\cite{Achasov:2000zd,Achasov:2003ed, Achasov:2016bfr,Achasov:2018ujw} and CMD-2~\cite{Akhmetshin:2004gw} detectors. 
\bsp
As first observed in~\cite{DAgostini:1993arp}, a naive treatment of normalization-type systematic uncertainties would lead to a bias in the fit. For the data sets in Table~\ref{tab:pigamma-data_sets}, the systematic uncertainties of~\cite{Achasov:2000zd,Akhmetshin:2004gw} are explicitly given in percentages  and therefore interpreted as normalization uncertainties.
Likewise, we assume that the systematic uncertainties of~\cite{Achasov:2003ed,Achasov:2016bfr, Achasov:2018ujw} can be attributed primarily to effects in the same category
and thus treat all the systematics uncertainties as $100\%$ correlated. Accordingly, we employ the iterative solution strategy introduced in~\cite{Ball:2009qv} to treat the normalization uncertainties in a consistent manner and  consider both fits with diagonal and full covariance matrices to better monitor the role of the correlations, in analogy to the strategy in~\cite{Hoferichter:2019gzf}.
\esp
%--------------------------------------------------------------------------------------------------------
\subsection{ Fits to SND}
%--------------------------------------------------------------------------------------------------------
\label{sec:pigamma-SND}
\begin{table}
	\centering
	\renewcommand{\arraystretch}{1.3}
	\begin{tabular}{lcc}
		\toprule
		& \multicolumn{1}{c}{diagonal} & \multicolumn{1}{c}{full}\\
                \midrule 
		$\chi^2/\text{dof}$ & $116.9/100$ & $151.3/100$ \\
		& $=1.17$ & $=1.51$ \\
		$p$-value & $0.12$ & $7\times 10^{-4}$  \\
		$\Mw \ [\text{MeV}]$ & $782.55(3)$ & $782.58(3)$ \\
		$\Gw \ [\text{MeV}]$ & $8.73(7)$ & $8.68(6)$ \\
		$\Mphi \ [\text{MeV}]$& $1019.18(5)$ & $1019.18(6)$ \\
		$\Gphi \ [\text{MeV}]$ & $4.24(16)$ & $4.27(17)$ \\
		$c_\omega \ [\text{GeV}^{-1}]$ & $2.95(2)$ & $2.95(3)$ \\
		$c_\phi \ [\text{GeV}^{-1}]$ & $-0.378(11)$ & $-0.382(13)$ \\
		$10^4\times \xi$ & $3.5(1.3)$ & $4.0(1.0)$ \\
		$10^{11}\times a_\mu^{\pi^0\gamma}|_{\leq \SI{1.35}{\GeV}}$ & $44.05(24)$ & $44.14(57)$\\ 
		\bottomrule
		\renewcommand{\arraystretch}{1.0}
	\end{tabular}
	\caption{Fits to the combined SND data sets~\cite{Achasov:2000zd,Achasov:2003ed, Achasov:2016bfr,Achasov:2018ujw}, for diagonal uncertainties and full covariance matrices. All errors refer to fit uncertainties only.}
	\label{tab:pigamma-fits_SND}
\end{table}

\bsp
First, we perform fits to the SND data sets~\cite{Achasov:2000zd,Achasov:2003ed, Achasov:2016bfr,Achasov:2018ujw}, with the results shown in Table~\ref{tab:pigamma-fits_SND}. We display the best $\chi^2$ results for both the diagonal fit and also the fully correlated one.  Only the fit uncertainties are displayed in Table~\ref{tab:pigamma-fits_SND} at this step, as we will add the systematic uncertainties of our approach later.    Fit errors are already inflated by the scale factor
\begin{equation}
\label{eq:pigamma-SF}
S=\sqrt{\chi^2/\text{dof}},
\end{equation}
to account for potential inconsistencies between the data sets following the PDG prescription~\cite{PDG:2020}. 

In contrast to~\cite{Hoferichter:2019gzf}, we do not include the  $\omega'(1420)$ or other excited vector mesons in the fits since their residues come out consistent with zero, in such a way that their inclusion does not improve the quality of the fit. This strategy is consistent with the 
observation of a negligible cross section above $\SI{1.4}{\GeV}$ in~\cite{Achasov:2018ujw}. Similarly, the data points above the $\phi$ region are scarce, so that additional free parameters in the conformal polynomial 
in the parameterization of $a(q^2)$ 
also do not improve the fits. Therefore, we will use the conformal polynomial to implement the chiral low-energy theorem $F_{3\pi}$ (with $S$-wave singularities removed), but do not add additional free parameters.  

The accuracy of the center-of-mass energy determination of the data set~\cite{Achasov:2003ed} is worse than the accuracy of the $\omega$ mass value. Therefore, an energy-scale bias $\Delta E$ was introduced in~\cite{Achasov:2003ed}. A separate fit to~\cite{Achasov:2003ed} indeed produces a smaller $ \omega$ mass that is not compatible with the most precise measurement~\cite{Achasov:2016bfr}. Therefore, we allow for an energy rescaling for~\cite{Achasov:2003ed},
\begin{equation}
\sqrt{s}\to\sqrt{s} +\xi (\sqrt{s}-M_{\pi^0}).
\end{equation}
The introduced scaling indeed leads to a considerable improvement of the fits, and its value around $\xi\sim 10^{-4}$ comes out in agreement with the energy-bias uncertainties. Similar rescalings within the quoted energy uncertainties were also found to improve the fit quality for the $2\pi$~\cite{Colangelo:2018mtw} and $3\pi$~\cite{Hoferichter:2019gzf} channels. In the case of $\pi^0\gamma$, the data set from~\cite{Achasov:2003ed} is the only one for which we see a need for such a rescaling. 

\begin{table}
	\centering
	\renewcommand{\arraystretch}{1.3}
	\begin{tabular}{lcc}
		\toprule
		& \multicolumn{1}{c}{diagonal} & \multicolumn{1}{c}{full}\\
                \midrule 
		$\chi^2/\text{dof}$ & $42.50/40$ & $57.39/40$ \\
		& $=1.06$ & $=1.43$ \\
		$p$-value & $0.36$ & $0.04$  \\
		$\Mw \ [\text{MeV}]$ & $782.53(14)$ & $782.68(9)$ \\
		$\Gw \ [\text{MeV}]$ & $8.25(28)$ & $8.41(19)$ \\
		$\Mphi \ [\text{MeV}]$& $1019.18(7)$ & $1019.18(6)$ \\
		$\Gphi \ [\text{MeV}]$ & $3.90(21)$ & $3.90(17)$ \\
		$c_\omega \ [\text{GeV}^{-1}]$ & $2.91(7)$ & $2.92(13)$ \\
		$c_\phi \ [\text{GeV}^{-1}]$ & $-0.342(13)$ & $-0.341(17)$ \\
		$10^{11}\times a_\mu^{\pi^0\gamma}|_{\leq \SI{1.35}{\GeV}}$ & $44.88(99)$ & $44.48(3.05)$\\ 
		\bottomrule
		\renewcommand{\arraystretch}{1.0}
	\end{tabular}
	\caption{Fits to the CMD-2 data set~\cite{Akhmetshin:2004gw}.}
	\label{tab:pigamma-fits_CMD-2}
\end{table}

We observe that the correlated fit produces larger uncertainties for the parameters and the HVP contribution compared to the diagonal one. Otherwise, the central values of the parameters of both fits are in good agreement within uncertainties. Besides, we find that the correlated fit has a worse description than the diagonal fit, which is a general observation of the iterative fit strategy~\cite{Ball:2009qv} concerning normalization uncertainties. In fact, this effect may be overestimated here because all systematic uncertainties of~\cite{Achasov:2003ed,Achasov:2016bfr, Achasov:2018ujw} were assumed to contribute in that category, so that the description could likely be improved if more details on the systematic uncertainties were available. At present, the relatively large $\chi^2$ of the correlated fit is mainly driven by~\cite{Achasov:2018ujw}: a fit to this data set alone gives a $\chi^2/\text{dof}=88.7/54=1.64$ and a $p$-value of $0.2\%$. The fact that the $p$-value drops by another factor of $3$ in the combined SND fit thus points to some minor tensions among~\cite{Achasov:2000zd,Achasov:2003ed, Achasov:2016bfr,Achasov:2018ujw}.   
\esp

%--------------------------------------------------------------------------------------------------------
\subsection{ Fits to CMD-2}
\label{sec:CMD2}
%--------------------------------------------------------------------------------------------------------
\label{sec:pigamma-CMD-2}

\begin{table}
	\centering
	\renewcommand{\arraystretch}{1.3}
	\begin{tabular}{lcc}
		\toprule
		& \multicolumn{1}{c}{diagonal} & \multicolumn{1}{c}{full}\\
                \midrule 
		$\chi^2/\text{dof}$ & $173.3/146$ & $238.6/146$ \\
		& $=1.19$ & $=1.63$ \\
		$p$-value & $0.06$ & $2\times 10^{-6}$  \\
		$\Mw \ [\text{MeV}]$ & $782.55(3)$ & $782.58(3)$ \\
		$\Gw \ [\text{MeV}]$ & $8.71(7)$ & $8.65(6)$ \\
		$\Mphi \ [\text{MeV}]$& $1019.20(4)$ & $1019.21(4)$ \\
		$\Gphi \ [\text{MeV}]$ & $4.08(13)$ & $4.07(13)$ \\
		$c_\omega \ [\text{GeV}^{-1}]$ & $2.95(2)$ & $2.93(3)$ \\
		$c_\phi \ [\text{GeV}^{-1}]$ & $-0.363(9)$ & $-0.358(10)$ \\
		$10^4\times \xi$ & $3.5(1.3)$ & $4.1(1.0)$ \\
		$10^{11}\times a_\mu^{\pi^0\gamma}|_{\leq \SI{1.35}{\GeV}}$ & $44.04(23)$ & $43.82(58)$\\ 
		\bottomrule
		\renewcommand{\arraystretch}{1.0}
	\end{tabular}
	\caption{Fits to the combined data sets as shown in Table~\ref{tab:pigamma-data_sets}.}
	\label{tab:pigamma-fits_com}
\end{table}

Next, we turn to the fits to the CMD-2 data~\cite{Akhmetshin:2004gw}. Although there is only a single data set, it covers almost the entire relevant energy region.  The results are given in Table~\ref{tab:pigamma-fits_CMD-2}, in the same form as the SND fits, the only exception being the exclusion of the rescaling parameter. For comparison, the fit uncertainties are also inflated by the scale factor~\eqref{eq:pigamma-SF}. 

As for the SND fits, we again find internal consistency for the parameters of the diagonal and the correlated fits. A minor difference concerns the mass and width of the $\omega$, which display relativity large upward shifts once the correlations are included.

\bsp
Even once accounting for VP corrections, see Sect.~\ref{sec:resonances}, our result for the $\omega$ mass is substantially smaller than in~\cite{Akhmetshin:2004gw}, which quotes $\barMw=783.20(13)(16)\MeV$. A key difference to our formalism is that the vector-meson-dominance ansatz from~\cite{Akhmetshin:2004gw} (see also~\cite{Achasov:1989mh})  permits a complex phase between the $\omega$ and $\rho$ contributions, which cannot be physical because it violates analyticity and unitarity, e.g., by introducing an imaginary part below the respective thresholds.
In our fits, we do not see a conflict with the $\omega$ mass extracted from $3\pi$ cross sections, and thus conclude that the result from~\cite{Akhmetshin:2004gw} is likely affected by the unphysical phase.  
\esp

Compared to the SND fits, we observe that the width of the $\phi$ comes out appreciably smaller, albeit with rather large fit uncertainties. 
This observation will also be reflected in the determination of the width of the $\phi$ in the combined fit presented in the next section.

\begin{figure}
	\centering
	\includegraphics[width=\linewidth]{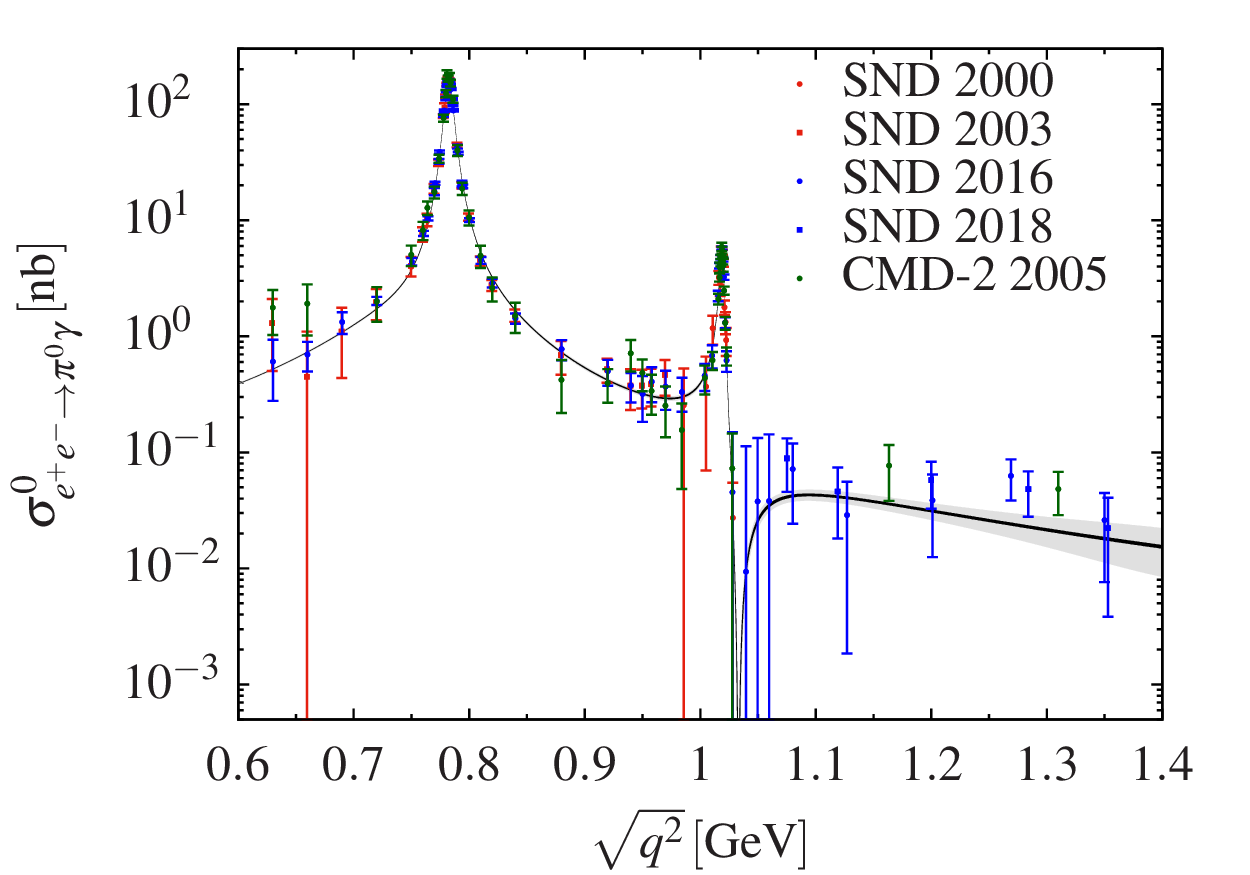}
	\caption{The final fit to the $e^+e^-\to\pi^0\gamma$ data sets as listed in Table~\ref{tab:pigamma-data_sets} (with VP removed everywhere), where the gray band indicates the full uncertainty and the black band indicates the fit uncertainty.} 
	\label{fig:pigamma-pi0gamma}
\end{figure}

%--------------------------------------------------------------------------------------------------------
\subsection{ Combined fits}
%--------------------------------------------------------------------------------------------------------

Finally, our combined SND and CMD-2 fit results are presented in Table~\ref{tab:pigamma-fits_com}, including all the data sets listed in Table~\ref{tab:pigamma-data_sets}.  We take the correlated full fit as our central value, and define our systematic uncertainties as the maximum deviations from the different fit variations discussed in Sect.~\ref{sec:TFF}. In all cases, the uncertainties are statistics dominated, in part because a main source of systematic uncertainty from the $3\pi$ channel~\cite{Hoferichter:2019gzf}, the degree of the conformal polynomial, does not become relevant here given that the observed cross section becomes negligibly small around $1.4\GeV$, with few data points above the $\phi$ resonance.

\begin{figure*}
	\centering
	\includegraphics[width=0.49\linewidth,clip]{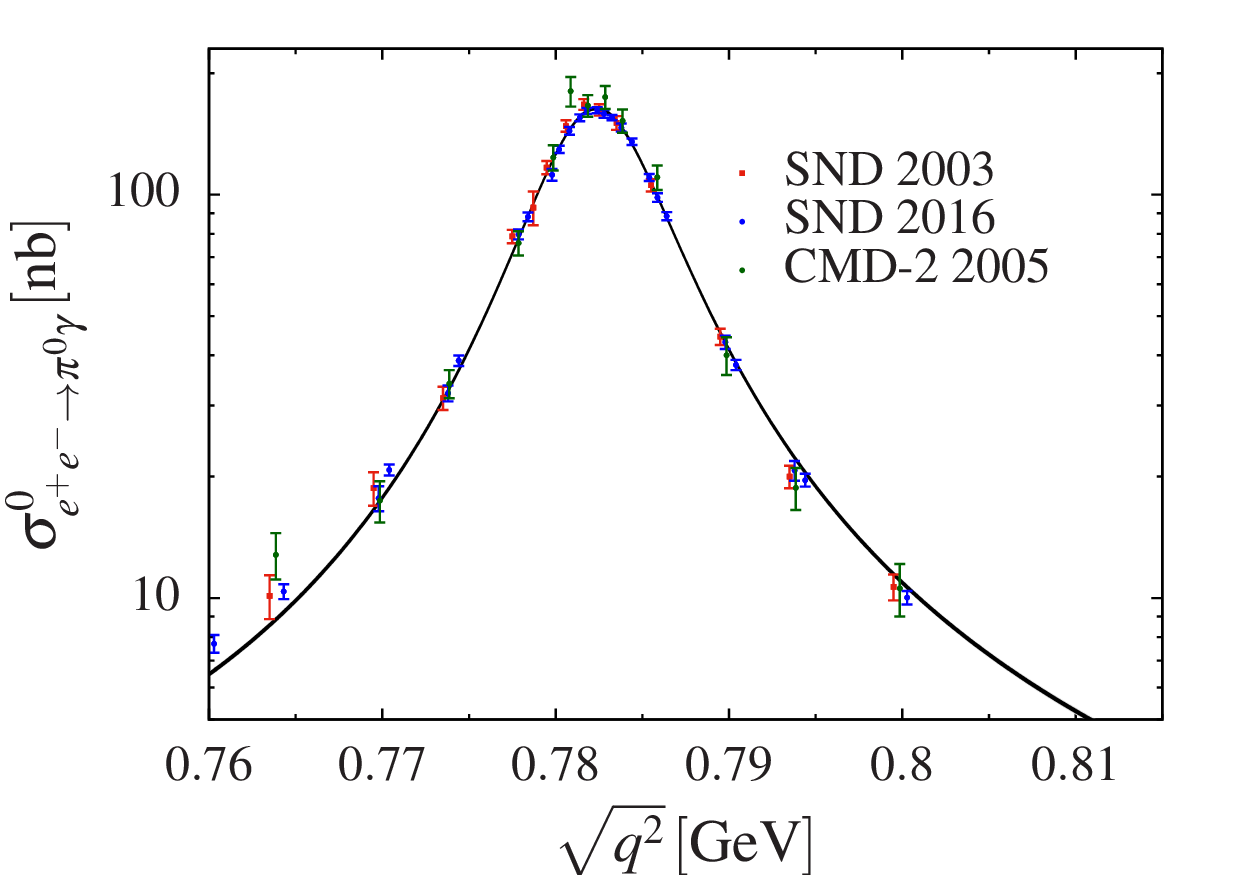}
	\includegraphics[width=0.49\linewidth,clip]{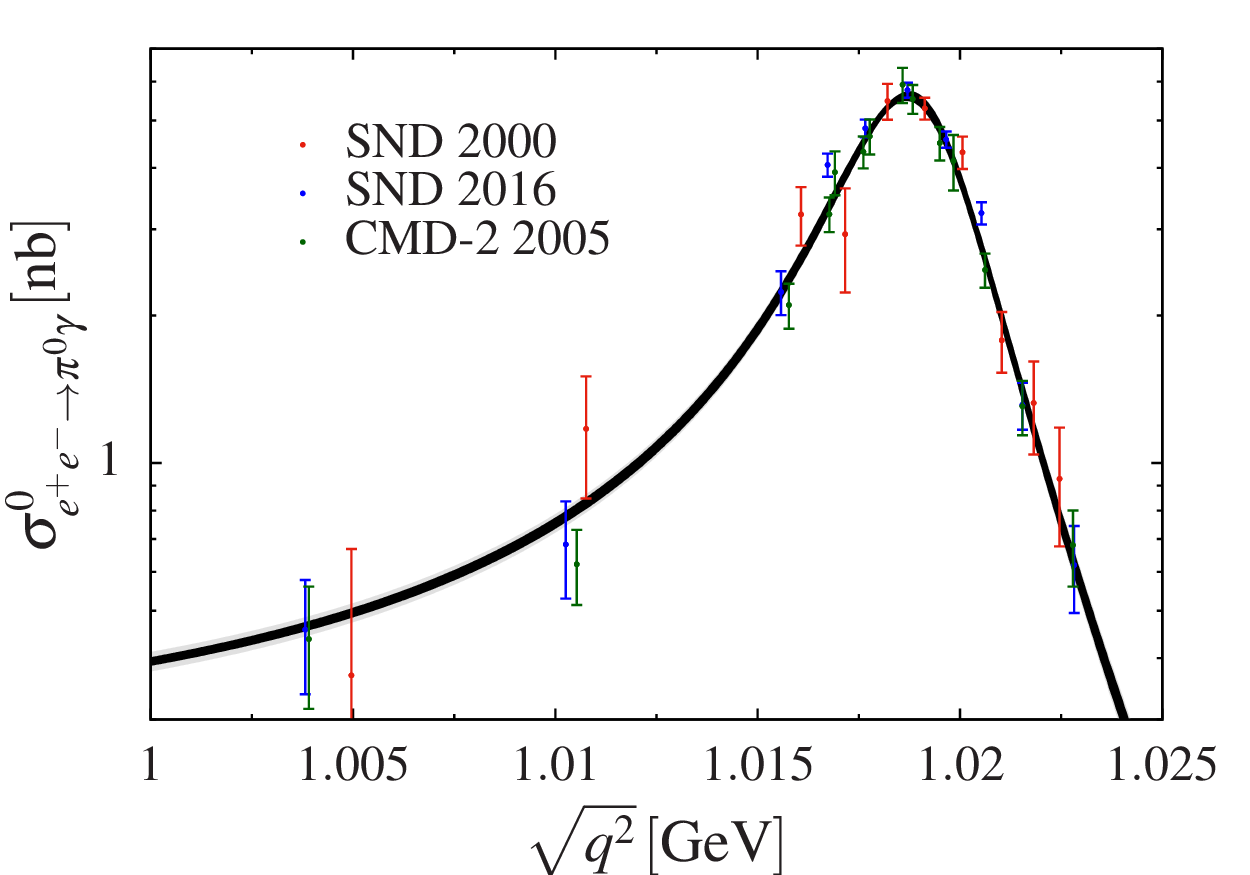}
	\caption{Fit around the $\omega$ and $\phi$ resonance regions. The black band represents the fit uncertainties, and the gray band indicates the total uncertainty. 
	}
	\label{fig:pigamma-cross_section2}
\end{figure*}

The combined fit, although dominated by the SND data, reflects some inconsistencies between SND and CMD-2. 
Most prominently, the downward shift of the width of the $\phi$ in comparison to Table~\ref{tab:pigamma-fits_SND} is due to the CMD-2 data~\cite{Akhmetshin:2004gw}. The coupling $c_\phi $ is also affected and shifted to a smaller value compared to the SND fits. 
Comparing the residues $c_\omega$ and $c_\phi$ to the $3\pi$ fit~\cite{Hoferichter:2019gzf}, $c_\omega=2.86(2)(4)$ and $c_\phi=-0.386(4)(2)$,
we observe reasonable agreement, which indeed is better for $c_\omega$ than for $c_\phi$. Taken together with the fact that also the $\phi$ width from the CMD-2 $\pi^0\gamma$ data drives the combined fit away from the $3\pi$ value, we conclude that indeed the interchannel consistency is better for the SND data sets. 
Figure~\ref{fig:pigamma-pi0gamma} illustrates our final preferred fit, with close-up views of the $\omega$ and $\phi$ regions in Fig.~\ref{fig:pigamma-cross_section2}. 

\bsp
The final result for the $\omega$ and $\phi$ parameters reads
\begin{align}
\label{eq:pigamma-omega_phi_fit}
\Mw&=782.58(3)(1)\MeV=782.58(3)\MeV,\notag\\
\Gw&=8.65(6)(1)\MeV=8.65(6)\MeV ,\notag\\
\Mphi&=1019.21(4)(3)\MeV=1019.21(5)\MeV,\notag\\
\Gphi&=4.07(13)(1)\MeV=4.07(13)\MeV,
\end{align}
with systematic errors in the second brackets derived as described above. 
We stress that these resonance parameters do not include VP corrections, see Sect.~\ref{sec:resonances} for a more detailed discussion. 
\esp

%-----------------------------------------------------------------------------
\section{Consequences for the anomalous magnetic moment of the muon}
\label{sec:amu}
%-----------------------------------------------------------------------------

\bsp
The HVP contribution to the anomalous magnetic moment of the muon reads~\cite{Bouchiat:1961,Brodsky:1967sr}
\beq
	a_\mu^\text{HVP} = \Big( \frac{\alpha m_\mu}{3\pi} \Big)^2
        \int_{s_\mathrm{thr}}^\infty \diff s \frac{\hat K(s)}{s^2}
        R_\mathrm{had}(s),
\label{eq:amuHVP}
\eeq
where the $R$-ratio 
\beq
	\label{eq:Rratio}
	R_\mathrm{had}(s) = \frac{3s}{4\pi\alpha^2}\sigma^{(0)}(e^+e^-\to\mathrm{hadrons}(+\gamma))
\eeq
is a substitute for the (bare) hadronic cross section and the kernel function $\hat K(s)$ is known analytically in terms of the center-of-mass energy $s$ and the muon mass $m_\mu$. By convention, the hadrons in the final state of the cross section include photons, so that the $\pi^0\gamma$ channel is actually the first to contribute and sets the integration threshold in~\eqref{eq:amuHVP} to $s_\mathrm{thr}=M_{\pi^0}^2$.
Based on the fits presented in the previous section our central result for the HVP contribution from the $\pi^0\gamma$  channel becomes
\begin{align}
\label{eq:pigamma_final}
a_\mu^{\pi^0\gamma}|_{\leq \SI{1.35}{\GeV}}=43.8(6)(1)\times 10^{-11}
=43.8(6)\times 10^{-11},
\end{align}
where the second uncertainty  is systematic.\footnote{We quote the HVP integral up to the last data point in~\cite{Achasov:2018ujw} that shows a nonvanishing cross section, and in the comparison to other work indicate the energy up to which the sum of exclusive channels is considered. However, in practice the energy region above $1.35\GeV$ can simply be ignored in the $\pi^0\gamma$ channel, see~\cite{Achasov:2018ujw}. An extrapolation of our results beyond $1.35\GeV$ suggests that this region contributes less than $0.1\times 10^{-11}$ to the HVP integral.} In comparison to the most recent direct-data-integration analyses, our result is in  good agreement  with $a_\mu^{\pi^0\gamma}|_{\leq \SI{1.8}{\GeV}}=44.1(1.0)\times 10^{-11}$~\cite{Davier:2019can},
with a slight improvement in the uncertainty thanks to the incorporation of the general QCD constraints.
 The small difference to $a_\mu^{\pi^0\gamma}|_{\leq \SI{1.937}{\GeV}}=45.8(1.0)\times 10^{-11}$~\cite{Keshavarzi:2019abf} partly originates from the application of the trapezoidal rule to scarce data in the tails of the $\omega$ resonance, similarly to the case of $3\pi$. Higher-order interpolations to the data combination of~\cite{Keshavarzi:2019abf} indeed move the HVP contribution towards~\eqref{eq:pigamma_final}. Our analysis does not support values as low as $a_\mu^{\pi^0\gamma}|_{\leq \SI{2.0}{\GeV}}=40.0(1.6)\times 10^{-11}$~\cite{Jegerlehner:2017gek}, which is based on a Breit--Wigner description of $\omega$ and $\phi$. The analysis~\cite{Davier:2019can} has updated~\cite{Davier:2017zfy} to account for  the threshold contribution $a_\mu^{\pi^0\gamma}|_{\leq \SI{0.6}{\GeV}}=1.2 \times 10^{-11}$, which was already included in~\cite{Keshavarzi:2018mgv,Keshavarzi:2019abf}.  It was determined in~\cite{Hagiwara:2003da} based upon a combination of the chiral-anomaly term and $\omega$-meson dominance~\cite{Achasov:2002bh}. This result is in line with our finding for the threshold region, $a_\mu^{\pi^0\gamma}|_{\leq \SI{0.6}{\GeV}}=1.3 \times 10^{-11}$.   Indeed, the agreement between the prediction and the cross section of the first few data points was already observed in~\cite{Hagiwara:2003da}. Although these small differences are negligible at the current level of accuracy required for HVP, it is reassuring that the dispersive analysis also corroborates current estimates for the $\pi^0\gamma$ channel, making significant changes in HVP in the energy region up to $1\GeV$ increasingly unlikely. 
 Other radiative effects beyond $\pi^0\gamma$, $\eta\gamma$, and the infrared-enhanced contributions in $\pi^+\pi^-\gamma$ are negligibly small compared to the current uncertainty of the full $a_\mu^\text{HVP}$, see, e.g.,~\cite{Moussallam:2013una}. 
\esp
%-----------------------------------------------------------------------------
\section{$\boldsymbol{\omega}$ and $\boldsymbol{\phi}$ resonance parameters}
\label{sec:resonances}
%-----------------------------------------------------------------------------

\begin{table}
	\centering
	\renewcommand{\arraystretch}{1.3}
	\begin{tabular}{lcccc}
		\toprule
		& $e^+e^-\to 3\pi$ & $e^+e^-\to \pi^0\gamma$ & combination\\
                \midrule 
		$\Mw \ [\text{MeV}]$ & $782.631(28)$ & $782.584(28)$ & $782.607(23)$\\
		$\Gw \ [\text{MeV}]$ & $8.71(6)$ & $8.65(6)$ & $8.69(4)$\\
		$\Mphi \ [\text{MeV}]$ & $1019.196(21)$ & $1019.205(55)$ & $1019.197(20)$\\
		$\Gphi \ [\text{MeV}]$ & $4.23(4)$ & $4.07(13)$ & $4.22(5)$\\
		\bottomrule
		\renewcommand{\arraystretch}{1.0}
	\end{tabular}
	\caption{$\omega$ and $\phi$ resonance parameters from $e^+e^-\to 3\pi$~\cite{Hoferichter:2019gzf}, $e^+e^-\to\pi^0\gamma$ (this work), and their combination. The final uncertainties for $\Mw$ and $\Gphi$ include a scale factor $S=1.2$. 
	All parameters do not include VP corrections, see Table~\ref{tab:PDG} for the comparison to the PDG parameters.}
	\label{tab:resonance}
\end{table}

\bsp
Our final results for the $\omega$ and $\phi$ resonance parameters as determined from $e^+e^-\to\pi^0\gamma$ are contrasted to the results from $e^+e^-\to3\pi$~\cite{Hoferichter:2019gzf} in Table~\ref{tab:resonance}. There is good agreement throughout, leading to the combination in the last column. Since the $\pi^0\gamma$ channel is statistics-dominated for all quantities, see~\eqref{eq:pigamma-omega_phi_fit}, the combination is straightforward despite the fact that the systematic errors related to the dispersive representation are correlated. Likewise, the statistical correlations among the resonance parameters (and with the residues) from the respective fits have a negligible impact on the combination.
$\Mw$ and $\Gphi$ require a small scale factor $S=1.2$ (defined in accordance with the PDG conventions~\cite{PDG:2020}). The slight tension for $\Gphi$ can be traced back to the CMD-2 data set~\cite{Akhmetshin:2004gw}, see Sect.~\ref{sec:CMD2}. However, we conclude that within uncertainties the $3\pi$ and $\pi^0\gamma$ channels yield a consistent picture for the $\omega$ and $\phi$ resonance parameters. 

To be able to compare our results to the PDG conventions, we need to restore the VP corrections that have been removed in the definition of the bare cross sections, which we will denote by a bar over the corresponding quantities. As argued in~\cite{Hoferichter:2019gzf}, this leads to the shifts
\begin{align}
 \barMw&=\bigg(1+\frac{e^2}{2\gwg^2}\bigg)\Mw=\Mw + 0.128(3)\MeV,\notag\\
 \barMphi&=\bigg(1+\frac{e^2}{2g_{\phi\gamma}^2}\bigg)\Mphi=\Mphi + 0.260(3)\MeV,
\end{align}
where the couplings are related to the respective $e^+e^-$ widths, e.g., $\Gamma_{\omega\to e^+e^-}=e^4\Mw/(12\pi \gwg^2)$, and the uncertainties have been propagated from the PDG values~\cite{PDG:2020} (with potential differences to our determinations being higher-order effects). While otherwise shifts in the widths are negligible, there is an effect enhanced by $\rho$--$\omega$ mixing
\begin{align}
\barGw&=\Gw+\frac{e^2}{2\gwg^2}\Gw+\frac{\Mw^2}{\Gamma_\rho-\Gw}\frac{e^2}{g_{\rho\gamma}^2}\bigg(\frac{e^2}{\gwg^2}-2\eps_\omega\bigg)\notag\\
&=\Gw-0.06(2)\MeV,
\end{align}
where we have assigned a generous uncertainty because the estimate relies on a narrow-resonance assumption for the $\rho$.
\esp

\begin{table}
	\centering
	\renewcommand{\arraystretch}{1.3}
	\begin{tabular}{lcc}
		\toprule
		& $e^+e^-\to 3\pi,\pi^0\gamma$ & PDG\\
                \midrule 
		$\barMw \ [\text{MeV}]$ & $782.736(24)$ & $782.65(12)$\\
		$\barGw \ [\text{MeV}]$ & $8.63(5)$ & $8.49(8)$\\
		$\barMphi \ [\text{MeV}]$ & $1019.457(20)$ & $1019.461(16)$\\
		$\barGphi \ [\text{MeV}]$ & $4.22(5)$ & $4.249(13)$\\
		\bottomrule
		\renewcommand{\arraystretch}{1.0}
	\end{tabular}
	\caption{Comparison of $\omega$ and $\phi$ resonance parameters from $e^+e^-\to 3\pi,\pi^0\gamma$ to the PDG values, including VP corrections.}
	\label{tab:PDG}
\end{table}

\bsp
The resulting parameters, in comparison to the PDG values, are shown in Table~\ref{tab:PDG}. First, one sees that the $\phi$ mass agrees perfectly, with competitive uncertainties. This is an important observation because it demonstrates consistency between $e^+e^-\to 3\pi,\pi^0\gamma$  and $e^+e^-\to \bar K K$. The latter includes the BaBar measurements~\cite{Lees:2013gzt,Lees:2014xsh}, which, in contrast to all data sets for $e^+e^-\to\pi^0\gamma$ considered in this work as well as all the $e^+e^-\to 3\pi$ data sets relevant for the $\omega$ and $\phi$ parameters, have not been taken in energy-scan mode (at the VEPP-2M collider), but using initial-state radiation. The $\phi$ width also agrees within uncertainties, but not at the level of accuracy that can be achieved in the $\bar K K$ channel.

For the $\omega$ mass, its PDG value is dominated by the weighted average of determinations from $e^+e^-\to3\pi$ ($\barMw=782.68(9)(4)\MeV$~\cite{Akhmetshin:2003zn}, $\barMw=782.79(8)(9)\MeV$~\cite{Achasov:2003ir}), $e^+e^-\to\pi^0\gamma$ ($\barMw=783.20(13)(16)\MeV$~\cite{Akhmetshin:2004gw}), and $\bar p p\to\omega\pi^0\pi^0$ ($\barMw=781.96(13)(17)\MeV$~\cite{Amsler:1993pr}), where the spread among these determinations drives the scale factor $S=1.9$ and thus an uncertainty much larger than we obtain from $e^+e^-\to 3\pi,\pi^0\gamma$.

As described in Sect.~\ref{sec:CMD2}, we believe that the large value for the $\omega$ mass determined from $e^+e^-\to\pi^0\gamma$ in~\cite{Akhmetshin:2004gw} originates from an unphysical phase in the vector-meson-dominance model used for the extraction. For the $\bar p p$ reaction, the uncertainties are more difficult to assess than in the $e^+e^-$ processes because the shape of the background processes is unknown and because the width of the $\omega$ signal, $\Gamma=38.1(3)\MeV$, is dominated by the experimental resolution and much larger than the intrinsic $\omega$ width. Energy scans in $e^+e^-\to3\pi,\pi^0\gamma$, for which the entire amplitude can be reconstructed from general principles and whose energy resolution lies well below the $\omega$ width, should thus yield a much more reliable probe of the $\omega$ resonance parameters.

The $\omega$ mass can also be extracted via $\rho$--$\omega$ mixing in $e^+e^-\to2\pi$, and it has been known for a while~\cite{Lees:2012cj} that without further constraints such fits prefer significantly smaller values for $\Mw$ than both the PDG average and our determination from $e^+e^-\to 3\pi,\pi^0\gamma$. This conclusion was recently confirmed in~\cite{Colangelo:2018mtw} within a dispersive approach, leading to $\Mw=781.68(10)\MeV$, in significant tension with Table~\ref{tab:resonance}. However, given the high accuracy required in the $e^+e^-\to 2\pi$ channel, additional imaginary parts from the radiative channels $\pi^0\gamma$, $\pi\pi\gamma$, etc.\ may actually become relevant~\cite{Kubis}. Before their impact is better understood, we would thus consider the mass determination from $e^+e^-\to 3\pi,\pi^0\gamma$ to be more reliable.

As for the $\omega$ width, our value is consistent with earlier determinations from the $3\pi$ channel
($\barGw=8.68(23)(10)\MeV$~\cite{Akhmetshin:2003zn}, $\barGw=8.68(4)(15)\MeV$~\cite{Achasov:2003ir}), but lies above the PDG average by $1.5\sigma$. 
This tension is partly driven by an extraction from the reaction $p d\to {^{3}\text{He}}\,\omega$ ($\barGw=8.2(3)\MeV$~\cite{Wurzinger:1994bi}), but mostly due to an earlier measurement of $e^+e^-\to 3\pi$ by the ND collaboration ($\barGw=8.4(1)\MeV$~\cite{Aulchenko:1987ba}). However, it should be noted that the error quoted in~\cite{Aulchenko:1987ba} is only statistical, while the modern data sets~\cite{Akhmetshin:2003zn,Achasov:2003ir} provide a complete error estimate. Moreover, without access to the original data for $e^+e^-\to 3\pi$ from~\cite{Aulchenko:1987ba} it is impossible to assess its weight in global fits to the data base~\cite{Hoferichter:2019gzf}. In such a situation we do not believe it is adequate to keep the ND measurement in the average for $\Gw$ and would therefore consider our determination from modern $e^+e^-\to 3\pi,\pi^0\gamma$ data sets to be more reliable than the current PDG average.
\esp
 
%-----------------------------------------------------------------------------
\section{Summary}
\label{sec:summary}
%-----------------------------------------------------------------------------

\bsp
We have studied the cross section for $e^+e^-\to\pi^0\gamma$ in a dispersive framework, which implements constraints from analyticity, unitarity, and crossing symmetry as well as low-energy theorems for 
the $\gamma\to3\pi$ amplitude and the transition form factor for $\pi^0\to\gamma\gamma^*$. The relation between this form factor and the $e^+e^-\to \pi^0\gamma$ cross section forms the basis for the subsequent data analysis. 

As the next step, we considered the full data sets for  $e^+e^-\to \pi^0\gamma$ from SND and CMD-2. An iterative fit algorithm was applied to eliminate the D'Agostini bias. Some  tensions among different data sets exist and the resulting scale factor of the global fit turns out to be larger compared to those of similar analyses of the $e^+e^-\to 2\pi$ and $e^+e^- \to 3\pi$ reactions, which in part can be traced back to assumptions  
necessary for the details of the systematic uncertainties. 
However, we did not find any data set that needed to be excluded because of severe tensions nor did we identify problematic outliers in the data sets. 

As a first application, we evaluated the $\pi^0\gamma$ contribution to HVP, with our central result given in~\eqref{eq:pigamma_final}. 
In general, the outcome is in good agreement with analyses using a direct integration of the data, with a slightly reduced uncertainty thanks to the global fit function defined by the dispersive representation. 
In combination with previous work on $e^+e^-\to 2\pi$ and $e^+e^- \to 3\pi$,
 the three largest channels below $\SI{1}{\GeV}$ 
 have now been subject to scrutiny using constraints from analyticity, unitarity, and low-energy theorems. 
 
Finally, we studied the resulting $\omega$ and $\phi$ resonance parameters first from $e^+e^-\to\pi^0\gamma$ and then in combination with $e^+e^-\to 3\pi$. Contrary to previous analyses, we find good agreement between the two channels, suggesting that a previous tension could be due to unphysical complex phases in a vector-meson-dominance model employed for the $e^+e^-\to\pi^0\gamma$ channel. Comparing the combined determinations to the current PDG averages, see Table~\ref{tab:PDG}, we observe that for the $\phi$ mass, the value obtained from $e^+e^-\to3\pi,\pi^0\gamma$ 
agrees perfectly at a similar level of precision, demonstrating consistency between extractions from $e^+e^-\to3\pi,\pi^0\gamma$ and $e^+e^-\to\bar K K$, the latter dominating the PDG average. The width also comes out consistent, but with larger uncertainty than from the $\bar K K$ channel.
For the $\omega$, we find that the combination of $e^+e^-\to3\pi$ and $e^+e^-\to\pi^0\gamma$ determines its mass at a level not far from the $\phi$ mass, and argue that the resulting values both for the $\omega$ mass and the width are more reliable than the current PDG averages. 
However, the tension with the $\omega$ mass determination from the $2\pi$ channel persists, suggesting that an improved understanding of isospin-breaking effects therein will become necessary.  
\esp

\begin{acknowledgements}
\bsp
We thank A.~Keshavarzi for helpful discussions and for sending us the $\pi^0\gamma$ data combination of~\cite{Keshavarzi:2018mgv,Keshavarzi:2019abf}. 
Financial support by the SNSF (Project No.\ PCEFP2\_181117)
 and the DFG (CRC 110,
``Symmetries and the Emergence of Structure in QCD'') is gratefully acknowledged.
\esp 
\end{acknowledgements}

\end{document}